# Diversifying portfolios of U.S. stocks with crude oil and natural gas: A regime-dependent optimization with several risk measures

Hayette GATFAOUI

Associate Professor, hgatfaoui@gmail.com, h.gatfaoui@ieseg.fr

IÉSEG School of Management (LEM – CNRS), Finance, Audit & Control Department, Paris Campus

Socle de la Grande Arche, 1 Parvis de La Défense, 92044 Paris La Défense ;

and Research associate at the University of Paris 1, Pantheon-Sorbonne

**Abstract:** Energy markets are strategic to governments and economic development. Several commodities compete as substitutable energy sources and energy diversifiers. Such competition reduces the energy vulnerability of countries as well as portfolios' risk exposure. Vulnerability results mainly from price trends and fluctuations, following supply and demand shocks. Such energy price uncertainty attracts many market participants in the energy commodity markets. First, energy producers and consumers hedge adverse price changes with energy derivatives. Second, financial market participants use commodities and commodity derivatives to diversify their conventional portfolios. For that reason, we consider the joint dependence between the United States (U.S.) natural gas, crude oil and stock markets. We use Gatfaoui's (2015) time-varying multivariate copula analysis and related variance regimes. Such approach handles structural changes in asset prices. In this light, we draw implications for portfolio optimization, when investors diversify their stock portfolios with natural gas and crude oil assets. We minimize the portfolio's variance, semi-variance and tail risk, in the presence and the absence of constraints on the portfolio's expected return and/or U.S. stock investment. The return constraint reduces the performance of the optimal portfolio. Moreover, the regime-specific portfolio optimization helps implement an enhanced active management strategy over the whole sample period. Under a return constraint, the semi-variance optimal portfolio offers the best risk-return tradeoff, whereas the tail-risk optimal portfolio offers the best tradeoff in the absence of a return constraint.

## 1. Introduction

Energy commodities are of interest to both the government and several parties acting in the financial markets. From the viewpoint of countries and governments, energy prices are crucial to economic development, and raise energy vulnerability concerns (e.g. energy shortages and costs). Energy price uncertainty, such as price upsurges, can severely impair energy firms as well as energy consumers, and therefore economic activity. From the viewpoint of financial markets, commodity and commodity derivatives markets are useful to energy producers and consumers as well as portfolio managers. They play an important role in related risk-sharing processes. As a result, energy commodities have become an asset class, which is widely used for diversification, hedging or speculation prospects (de Roon et al., 2000; Gorton and Rouwenhorst, 2006; Jin and Jorion, 2006). According to the United States (U.S.) Commodity Futures Trading Commission,[1] the notional value of commodity index investments has evolved from $39.6/$11.4 Billion on December 31$^{st}$, 2007 to $33.9/$14.3 Billion on January 31$^{st}$, 2013 with respect to WTI crude oil and natural gas markets.[2] Moreover, energy and financial markets are known to interact to a large extent, and exhibit a joint dependency (Barsky and Kilian, 2004; Hamilton, 1985; Kilian, 2009).

The proposed research tackles the interaction between the U.S. stock, natural gas, and crude oil markets as well as its significance to portfolio management. For this purpose, we consider daily natural gas and crude oil prices as well as daily quotes of the Standard and Poor's 500 index. We build our study on the findings of Gatfaoui (2015). The author studies the three-dimensional dependence structure between the U.S. stock, crude oil, and natural gas markets. Findings highlight structural changes/breaks over time. Hence, the resulting multivariate dependence structure is time-varying, and exhibits several variance regimes. Our added value and contributions extend previous study to portfolio management. We scrutinize practical implications for portfolio optimization across variance regimes. For this purpose, we consider several risk measures and various minimization constraints. The risk measures under consideration consist of the variance, semi-variance (e.g. downside variance) and tail risk (i.e. probability of underperforming a specified return threshold). We exploit the diversification power of energy commodities while minimizing the portfolio's (loss) risk. The employed risk measures incorporate the joint dependence structures of stock and energy commodity markets across regimes. Such regime-specific analysis allows for implementing an active portfolio management strategy. An efficient strategy should rely on tail risk minimization in the absence of a constraint on the portfolio's expected return. Differently, it should rely on semi-variance minimization in the presence of a return constraint.

The remainder of the paper is organized as follows. Section 2 summarizes the literature review. Section 3 introduces stock and energy market data, and recalls data attributes. This section displays key statistics and regime-specific multivariate dependencies. Section 4 applies previous results to portfolio optimization under various risk management schemes. We focus on stock portfolios, which are diversified with holdings in both natural gas and crude oil assets. A performance diagnostic scrutinizes the risk and return tradeoff of corresponding optimal portfolios. Finally, section 6 concludes and proposes possible future extensions.

---

[1] http://www.cftc.gov

[2] The notional value has recently fallen to $27/$8.9 Billion on June 30$^{th}$, 2015.

## 2. Literature review

We introduce acknowledged links between the U.S. stock market, crude oil and natural gas. We also highlight the recent role of energy commodities as portfolio diversifiers.

Index commodity trading favors the integration of stock and commodity markets (Domanski and Heath, 2007; Silvennoinen and Thorp, 2013; Tang and Xiong, 2012). It also enforces comovements between stock and commodity prices. Such pattern supports the reported price/return correlation between these two asset classes, and within the commodity asset class (Eckaus, 2008; Falkowski, 2011; Kilian, 2009; Kilian and Park, 2009; Parsons, 2010; Pindyck, 2004).[3] However, reported relationships are time-varying (Aloui et al., 2014; Brigida, 2014; Gatfaoui, 2015). Crude oil and natural gas prices are cointegrated so that they exhibit a long-run equilibrium from which they depart in the short run (Brigida, 2014; Hartley and Medlock, 2014; Miller and Ratti, 2009; Villar and Joutz, 2006). Crude oil and natural gas prices do indeed decouple in the short term whereas they couple in the longer term (Brown and Yücel, 2008; Gatfaoui, 2016; Hartley and Medlock, 2014; Ramberg and Parsons, 2012). As a result, crude oil-specific shocks need more or less time to contaminate natural gas prices. Such feature opens the door to potential market-timing strategies on the commodity market. Hence, incorporating crude oil and natural gas commodities into an investment portfolio can make sense. And, commodities can help mitigate the resulting portfolio's risk.

As regards risk mitigation prospects, commodity markets present some interests to market participants. Nowadays, their role as efficient portfolio diversifiers is strongly acknowledged (Goergiev, 2001; Jensen et al., 2000; Satyanarayan and Varangis, 1996). For example, futures on commodities help diversify equity portfolios. They contribute to reduce risk, and potentially improve portfolios' returns (Hensel and Ankrim, 1993; Lee and Leuthold, 1985). Incidentally, commodity markets exhibit a negative (and close to zero) correlation with stock markets (Harvey and Erb, 2006). Such negative or null correlation provides equity portfolios with diversification rewards (Conover et al., 2010). Commodities also contribute to improve the expected return of any conventional investment portfolio, which comprises stocks, bonds and cash (Bekkers et al., 2009). Thus, adding commodities to investment portfolios contributes to decrease (stabilize) portfolio volatility, and to stabilize (increase) corresponding returns (Conover et al., 2010; Christopherson et al., 2004; Harvey and Erb, 2006; Kim et al., 2011). In particular, they contribute to decrease the loss risk of portfolios (Daigler et al., 2016).

---

[3] For example, the speculative bubble has engendered oil price spikes during the financial crisis.

Including commodities into investment portfolios provides further benefits, specifically when equity markets are resilient and inflation is large. Commodity futures are negatively correlated with equity/bond returns, whereas they are positively correlated with changes in both expected inflation and unexpected inflation (Zvi and Rosanky, 1980; Gorton and Rouwenhorst, 2006). Thus, commodities offer a hedge against unexpected inflation (at least in the long run) because they are real assets (Greer, 2006; Hensel and Ankrim, 1993; Kaplan and Lummer, 1997). However, the diversification benefits of commodities vary over time. The diversification's effectiveness is sensitive to the financial market's disturbances (i.e. stressed market times; Büyükşahin et al., 2010; Cheung and Miu, 2010). The business cycle also influences such relationship (Chevallier and Ielpo, 2013; Chevallier et al., 2014). For example, the diversification power of commodities vanishes when the correlation between stock and commodity markets is strong (Silvennoinen and Thorp, 2013).

## 3. Energy and stock market data

We summarize the statistical features and key attributes of crude oil, natural gas, and Standard and Poor's index time series.

### 3.1. Dataset and properties

Analogously to Gatfaoui (2015), we consider daily logarithmic returns from January 8, 1997 to January 29, 2013 (i.e. 4032 observations per series). The data consist of Henry Hub Gulf Coast Natural Gas Spot Price (Gas) from the U.S. Energy Information Administration, the WTI crude oil Fixed Order Book price (Oil) from West Texas Intermediate exchange, and Standard and Poor's 500 index close (SP500) from Datastream. The selected sample window encompasses several market disturbances. Disturbances refer to the 1997 Asian crisis, the 1998 LTCM hedge fund default, the 2000-2002 dotcom bubble, the May 2005 U.S. credit crisis, the 2006 Amaranth hedge fund collapse, and the 2007-2009 subprime mortgage market crisis. Such disturbances impact the causal connections and dependencies between commodity and stock markets. And, they can potentially trigger structural changes in such relationships.

Gatfaoui (2015) reports skewed, leptokurtic, non-Gaussian, but mean-stationary returns. Moreover, crude oil, natural gas and SP500 index returns exhibit structural changes in their respective variances, while their respective means exhibit a unique regime. The reported five breakpoints allow for identifying six variance periods. The six periods summarize into four possible variance regimes (see Figure 1). The four variances regimes consist of a very low (VL), low (L), medium (M) and high (H) regime. Variance regimes are specific to each return series (see Table 1).

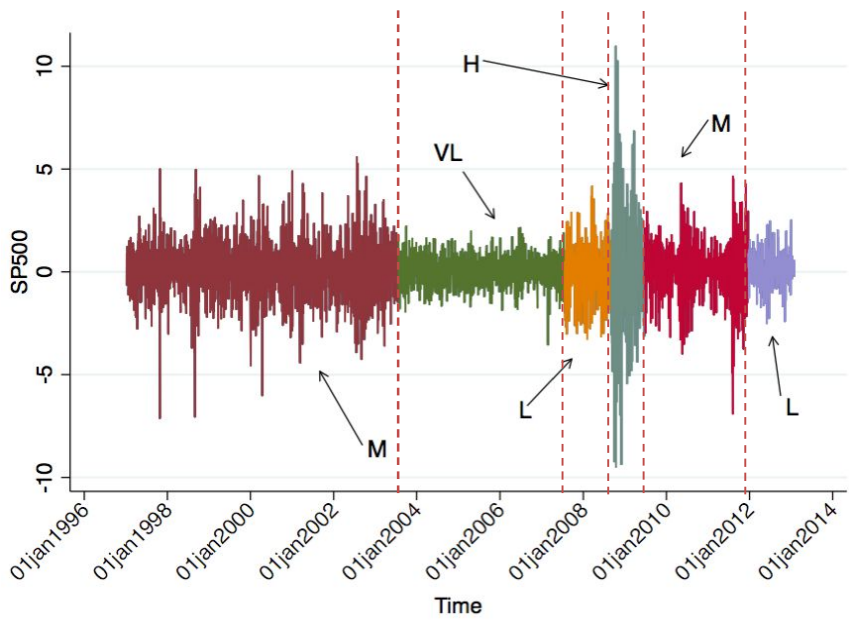

**Figure 1:** Variance regimes of SP500 index returns

**Table 1:** Variance regimes of Oil and Gas price returns, and SP500 index returns

| Period | Gas | | Oil | | SP500 | |
|---|---|---|---|---|---|---|
| | Dates* | Regime** | Dates | Regime | Dates | Regime |
| 1 | 01/08/97 10/02/01 | L | 01/08/97 01/23/98 | VL | 01/08/97 07/25/03 | M |
| 2 | 10/03/01 03/14/03 | H | 01/26/98 06/26/03 | M | 07/28/03 07/19/07 | VL |
| 3 | 03/17/03 02/20/07 | M | 06/27/03 08/20/08 | L | 07/20/07 08/29/08 | L |
| 4 | 02/21/07 03/19/09 | L | 08/21/08 06/10/09 | H | 09/02/08 06/22/09 | H |
| 5 | 03/20/09 01/11/10 | H | 06/11/09 10/27/11 | L | 06/23/09 12/20/11 | M |
| 6 | 01/12/10 01/29/13 | VL | 10/28/11 01/29/13 | VL | 12/21/11 01/29/13 | L |

* Start and end of a reference period.

** Variance regime of returns: very low (VL), low (L), medium (M), and high (H).

*3.2. Multivariate dependence structures across variance regimes*

At the multivariate level, previous return-specific variance regimes summarize into ten periods. These ten periods summarize into nine variance regimes (see Gatfaoui, 2015). The joint dependence structure between crude oil, natural gas and SP500 returns is investigated across these ten periods, or equivalently, across these nine variance regimes. The set of possible representations consist of the Clayton, Frank, Gauss, Gumbel and Student T copulas. Such copulas handle various tail behaviors (i.e. non existing, symmetric, lower or upper tail dependence; Cherubini *et al.*, 2004; Embrechts *et al.*, 2003; Gatfaoui, 2010; McNeil *et al.*, 2005; Patton, 2009; Sklar, 1973). As a result, the three-dimensional dependence structures investigate extremes' dependence (see Table 2).

**Table 2:** Optimal dependence structures across variance regimes

| Period | Start-End | Regime* SP500/Oil/Gas | Copula | Tail | Sign** |
|---|---|---|---|---|---|
| 1 | 01/08/97-01/23/98 | M/VL/L | Gumbel | Upper | + |
| 2 | 01/26/98-10/02/01 | M/M/L | Gumbel | Upper | + |
| **3** | 10/03/01-03/14/03 | **M/M/H** | Clayton | Lower | + |
| 4 | 03/17/03-02/20/07 | VL/L/M | Gauss | None | $\rho(SP500,Oil) = -$ <br> $\rho(SP500, Gas) = -$ <br> $\rho(Oil,Gas) = +$ |
| 5 | 02/21/07-07/19/07 | VL/L/L | Gumbel | Upper | + |
| 6 | 07/20/07-08/20/08 | L/L/L | Frank | None | + |
| 7 | 08/21/08-03/19/09 | H/H/L | Gauss | None | $\rho(SP500,Oil) = +$ <br> $\rho(SP500, Gas) = +$ <br> $\rho(Oil,Gas) = -$ |
| **8** | 03/20/09-01/11/10 | **M/M/H** | Gauss | None | + |
| 9 | 01/12/10-10/27/11 | M/L/VL | T | Symmetric | + |
| 10 | 10/28/11-01/29/13 | L/VL/VL | T | Symmetric | $\rho(SP500,Oil) = +$ <br> $\rho(SP500,Gas) = -$ <br> $\rho(Oil,Gas) = -$ |

* Variance regime of returns: very low (VL), low (L), medium (M) and high (H).
** Sign of correlation parameter(s), and $\rho(X,Y)$ represents the correlation between X and Y when a 3×3 correlation matrix is computed, and exhibits various correlation signs.

During the first three periods, SP500 variance regime drives tail dependence. Over the remaining periods, the variance regimes of energy commodities drive the dependence structures. The stock market disturbances of late 90s and early 2000s coincide with upper tail dependence. Tail dependency challenges portfolio management because it refers to comovements between the low and/or high extreme returns of SP500 index, crude oil and natural gas. According to Gatfaoui (2015), the balance between the variance regimes of crude oil, natural gas and SP500 drives tail dependence over time. As a result, the joint dependence structure between the three assets is regime-switching. Hence, a diversified portfolio composed of SP500 index, crude oil and natural gas assets needs to get reallocated across regimes. Such feature favors an active and regime-specific portfolio management since portfolio rebalancing should operate over each regime.

Given the diversifying role of commodities, we will consider a stock portfolio, which we immunize with energy commodities, such as crude oil and natural gas. Such commodity investment allows for benefiting from a direct exposure to commodity price changes. Moreover, including crude oil and natural gas commodities benefits from the short-term decoupling and long-term coupling of corresponding prices. It allows for capturing the market-timing strategies, which result from the regime-switching nature of stock, crude oil and natural gas markets.

4. **Portfolio optimization**

The dynamic nature of natural gas, crude oil and U.S. stock markets' dependencies causes shifts in their joint risk across reported regimes. In particular, interdependencies between U.S. commodity and stock markets drive the risk-sharing process, such as volatility spillovers. Such feature has significant implications for risk mitigation and portfolio management (i.e. dynamic asset allocation process and related optimization practice; Buckley et al., 2008). The portfolio's risk exposure and performance are altered because the time variation in risk needs to be reckoned. As an example, we consider a portfolio composed of U.S. stocks, which are diversified with holdings in both natural gas and crude oil assets. The investment portfolio under consideration comprises the SP500 index as well as natural gas and crude oil commodities. Such portfolio strategy exploits the hedging role of energy commodities against the liquidity risk of equity markets, among others. The optimal portfolio composition, or equivalently, the portfolio allocation profile is determined while minimizing the portfolio's risk exposure (e.g. limiting the risk of underperformance) over a given investment horizon. The investment horizon consists of a specific variance regime. In this light, we consider several risk measures, among which the variance of the portfolio's return, and the portfolio's downside risk such as semi-variance and tail risk (Mansini et al., 2014; Kolm et al., 2014). Corresponding portfolio optimization exploits previous copula representations as a robust tool (Kakouris and Rustem, 2014). In particular, we handle the asymmetry in returns, which is a significant issue to both portfolio selection (Harvey and Siddique, 2000; Kraus and Litzenberger, 1976; Samuelson, 1970), and portfolio decision making (Barberis and Huang, 2008; Mitton and Vorkink, 2007).

## 4.1. Variance minimization

We consider an investor whose wealth is devoted to a portfolio composed of the SP500, crude oil and natural gas assets. We label $w_1$, $w_2$ and $w_3$ those parts of wealth, which are respectively invested in each of the three assets. As formerly introduced by Markowitz (1952, 1959), the investor seeks a mean-variance efficient portfolio. Under a given return target, which represents the desired average portfolio performance over the investment horizon, the efficient portfolio's risk, which is measured by its return's variance, is minimized. Assuming that the investor dedicates all of his/her wealth to the portfolio, the portfolio's weights sum to unity. Hence, the investor needs to solve for the following optimization problem in order to determine his/her optimal portfolio (i.e. the best portfolio's allocation, subject to the performance constraint):

$$\min Var(R_P)$$
$$s.t. \begin{cases} E[R_P] = r \\ w_1 + w_2 + w_3 = 1 \end{cases} \quad (1)$$

where $R_P = w_1 R_{SP500} + w_2 R_{Oil} + w_3 R_{Gas}$ is the portfolio's return, and $E[R_P] = w_1 E[R_{SP500}] + w_2 E[R_{Oil}] + w_3 E[R_{Gas}]$ is its average counterpart ($E[.]$ is the expectation operator), $r$ is the targeted average return, and $Var(R_P)$ is the variance of the portfolio's return. When weights are positive (and, specifically, between 0 and 1), no short sale is allowed, while short selling is allowed under negative weights. When returns follow a Gaussian probability distribution, a simple solution to the optimization problem exists (Lintner, 1965; Markowitz, 1952; Sharpe, 1964). But, asset returns are far from being Gaussian because of their tail fatness and asymmetry properties. Moreover, their dynamic joint behaviors across reported regimes often deviate from the Gaussian setting.

Let $w$ and $R$ be the vectors of weights and asset returns respectively. The portfolio's return rewrites as $R_P = w' R$ where $w' = (w_1\ w_2\ w_3)$ is the transpose of vector $w$. Thus, the variance of the portfolio's return depends on the variance of the return vector $R$, and the joint dependence structure of the three portfolio constituents. As a result, the variance of the portfolio's return can rewrite as a function of the optimal copula, which describes the joint dependence structure of SP500, crude oil and natural gas returns. And, we can solve for optimization problem (1) while rewriting the components of the portfolio's variance $Var(R_P) = E[R_P{}^2] - (E[R_P])^2$ as follows:

$$E\left[R_P^2\right] = \iiint\limits_{[0,1]^3} \left(w_1 F_{SP500}^{-1}(u) + w_2 F_{Oil}^{-1}(v) + w_3 F_{Gas}^{-1}(w)\right)^2 c(u,v,w)\, du\, dv\, dw \quad (2)$$

and

$$E[R_P] = \iiint_{[0,1]^3} \left(w_1 F_{SP500}^{-1}(u) + w_2 F_{Oil}^{-1}(v) + w_3 F_{Gas}^{-1}(w)\right) c(u,v,w)\, du\, dv\, dw \qquad (3)$$

where $F_R(.)$ is the empirical cumulative distribution function of random variable $R$ (i.e. a given return series; see Deheuvels, 1979), and $F_R^{-1}(.)$ is its inverse counterpart, and $c(u,v,w)$ is the relevant copula density function (i.e. optimal copula representation). By so doing, we handle potential departures from normality, when computing the variance risk measure. We consider four possible cases while solving for the optimization problem above-mentioned (see Table 3). Such cases are compared to a benchmark portfolio, which consists of a naïve portfolio, or equivalently, an equally-weighted portfolio.[4] First, cases 1 and 2 assume a fixed target return (i.e. the investor's objective of average portfolio performance), whereas cases 3 and 4 let the optimization procedure determine the target return. The two latter cases are useful to study the feasibility of the optimization problem, and help correct for a too ambitious, or unrealistic target of portfolio performance. Under cases 1 and 2, the target return is equal to the naïve portfolio's performance plus a proportion of its absolute performance. In particular, the target return is set to the naïve portfolio's average return plus 0.5 times (i.e. fifty percent of) the absolute value of the naïve portfolio's average return. Hence, the optimal portfolio is required to outperform the naïve portfolio. Under cases 1 and 2, the optimization process yields the optimal portfolio's weights under a fixed target return. As an extension, cases 3 and 4 yield both the optimal portfolio's weights and corresponding optimal target return. The latter optimization cases help study the portfolio's risk-return tradeoff under the chosen risk measure.

**Table 3:** Optimization constraints and portfolio parameters

| Case # | Minimize subject to (un)constrained weights and target return | | | |
|---|---|---|---|---|
| 1 | $w_1$ is free | $w_2$ is free | $w_3$ is free | $r$ is fixed |
| 2 | $w_1$ lies in [0,1] | $w_2$ is free | $w_3$ is free | $r$ is fixed |
| 3 | $w_1$ is free | $w_2$ is free | $w_3$ is free | $r$ is free |
| 4 | $w_1$ lies in [0,1] | $w_2$ is free | $w_3$ is free | $r$ is free |

Second, cases 2 and 4 constrain the investment ($w_1$) in SP500 to lie between 0 and 1, while $w_2$ and $w_3$ weights (i.e. commodity investments) are unconstrained. Under such setting, the investor necessarily holds the SP500 index in his/her portfolio, and diversifies the index with (short/long) positions in commodity investments. Allowing negative values for $w_2$ and $w_3$ weights exploits the hedging effectiveness of commodities with respect to SP500. Specifically, diversification possibilities vanish in the presence of positive dependencies, so that the portfolio requires short sales to hedge against such positive interlinkages (particularly, when the desired target return is high).

---

[4] The naïve portfolio is an equally-weighted portfolio, which comprises SP500 index, and crude oil and natural gas commodities (i.e. each weight is equal to one-third).

We also introduce the upside potential ratio (UPR) as a performance indicator, which writes as UPR = $E[\max(R_P-r,0)] / (E[\min(R_P-r,0)^2])^{1/2}$. It is the ratio of the upper partial moment of order one to the square root of the lower partial moment of order 2. Over given investment horizon, the UPR is the ratio of all excess gains (with respect to the benchmark return *r*) to the downside risk (Sortino et al. 1999). In the UPR, lower partial moments measure the downside risk (Jarrow and Zhao, 2006; Unser, 2000).

*4.2. Minimizing downside risk: Semi-variance and tail risk*

So far, portfolio managers worry more about the downside risk (i.e. risk of negative returns, or risk of underperforming a return target) than the symmetric risk measure proposed by the variance. In this light, the square root of the variance, or equivalently, the standard deviation is as an absolute risk measure, which lacks information or risk representativeness in the presence of asymmetric returns. As an extension, several downside risk measures such as the semi-variance and tail risk have been proposed (Alexander, 1998; Steinbach, 2001). Downside risk measures help set up an optimal hedge, which offers protection against exposures to the risk of loss (Conlon and Cotter, 2013). The semi-variance focuses on the portfolio's returns, which underperform a fixed target return *r*. These underperforming returns lie below the targeted performance *r* (e.g. a benchmark return, which illustrates the minimum acceptable average return; Markowitz, 1959). In practice, the semi-variance measures the variance of the negative values of the difference ($R_P$-*r*) between the portfolio's return and its targeted performance. In this light, the semi-variance symbolizes a measure of regret because it focuses on the failure to reach the target return. Under such setting, the portfolio optimization process applies to a new risk measure, which consists of the semi-variance. The optimal portfolio's composition is thus determined while minimizing the downside risk (Lari-Lavassani and Li, 2003).

*Semi-variance optimization:*

Assuming the targeted portfolio's performance to be *r*, the semi-variance writes as *SemiVar ($R_P$)* = *Var[Min(0, $R_P$-r)]* and the new optimization problem rewrites:

$$\min SemiVar(R_P) = \min E\left[\{Min(0, R_P - r)\}^2\right] - \left(E\left[Min(0, R_P - r)\right]\right)^2$$

$$s.t. \begin{cases} E[R_P] = r \\ w_1 + w_2 + w_3 = 1 \end{cases} \quad (4)$$

with

$$E\left[\{Min(0, R_P - r)\}^2\right] = \iiint_{[0,1]^3} \{Min(0, w_1 F_{SP500}^{-1}(u) + w_2 F_{Oil}^{-1}(v) + w_3 F_{Gas}^{-1}(w) - r)\}^2 c(u,v,w) \, du \, dv \, dw \quad (5)$$

and

$$E\left[Min(0, R_P - r)\right] = \iiint_{[0,1]^3} Min(0, w_1 F_{SP500}^{-1}(u) + w_2 F_{Oil}^{-1}(v) + w_3 F_{Gas}^{-1}(w) - r) c(u,v,w) \, du \, dv \, dw \quad (6)$$

where $F_R(.)$ is the empirical cumulative distribution function of random variable $R$ (i.e. a given return series; see Deheuvels, 1979), and $F_R^{-1}(.)$ is its inverse counterpart, and $c(u,v,w)$ is the relevant copula density function (i.e. optimal copula representation). By so doing, we handle unfavorable deviations of the optimal portfolio's return from the targeted performance level, while accounting for empirical return behaviors.

*Tail risk optimization:*

The tail risk focuses on the probability of underperforming a given return threshold (e.g. worst-case study or analysis of bad scenarios; Dowd and Blake, 2006). Under such setting, the investor is risk averse, and hates losses, specifically extreme losses. He/she focuses on more or less extreme scenarios, under which his/her portfolio performs very poorly. In particular, the optimal portfolio allocation seeks to minimize the likelihood that the portfolio's return underperforms the chosen threshold (i.e. minimize the probability of a more or less extreme bad scenario). We focus on extreme bad scenarios, under which the bad return target (i.e. worst envisaged critical threshold) consists of the five percent quantile of SP500 index return. We label such quantile $q_{5\%}$. Under this setting, the investor seeks to minimize the probability that the portfolio's return belongs to the lower tail of SP500 returns' distribution (i.e. the five percent lowest values of SP500 returns) over the investment horizon. Such view requires the optimal portfolio to outperform SP500 index during disturbed market times, so that energy market investments hedge extreme exposures to stock market risk (i.e. extreme negative SP500 returns). Assuming the targeted portfolio's performance to be $r$, tail risk writes as $\lambda(q_{5\%}) = Pr(R_P \leq q_{5\%})$, and the optimization problem rewrites:

$$\min \lambda(q_{5\%}) = \min Pr(R_P \leq q_{5\%})$$
$$s.t. \begin{cases} E[R_P] = r \\ w_1 + w_2 + w_3 = 1 \end{cases} \quad (7)$$

with

$$\lambda(q_{5\%}) = \iiint_I c(u,v,w)\,du\,dv\,dw \qquad (8)$$

where Pr(.) is the probability operator, $I = \left\{(u,v,w) \in [0,1]^3 \,\middle|\, w_1 F_{SP500}^{-1}(u) + w_2 F_{Oil}^{-1}(v) + w_3 F_{Gas}^{-1}(w) \leq q_{5\%}\right\}$, $F_R(.)$ is the empirical cumulative distribution function of random variable $R$ (i.e. a given return series; see Deheuvels, 1979), and $F_R^{-1}(.)$ is its inverse counterpart, and $c(u,v,w)$ is the relevant copula density function (i.e. optimal copula representation). By so doing, we handle the optimal portfolio's downside risk because we target a reduced probability of unfavorable scenarios $\lambda(q_{5\%}) = Pr(R_P - q_{5\%} \leq 0)$, and we account for empirical return behaviors.

## 5. Optimal portfolios' attributes and performance

We introduce the results of portfolio optimization, among which portfolio allocations, optimal portfolios' performance, and related performance diagnostics. All optimization processes employ the Davidon-Fletcher-Powell method, with a $10^{-4}$ accuracy of gradient calculations (Davidon, 1991; Fletcher and Powell, 1963).

### 5.1. Variance optimization results

Table 5 displays all the optimization results while Table 4 displays relevant results for the naïve portfolio (i.e. benchmark portfolio).

**Table 4:** Annualized attributes of the naïve portfolio

| Regime | Average return | Target return $r$ | Standard Deviation | Skewness | Excess kurtosis | UPR |
|---|---|---|---|---|---|---|
| 1 | -23.9468 | -12.7882 | 22.5418 | -0.8306 | -2.6854 | 43.9470 |
| 2 | 2.8345 | 4.2816 | 28.0487 | 0.4609 | -1.6874 | 53.1637 |
| 3 | 33.8661 | 54.8640 | 43.5152 | 0.3409 | 0.5141 | 44.8949 |
| 4 | 12.8018 | 19.8022 | 28.6932 | 0.1175 | -0.7974 | 50.7773 |
| 5 | 17.9017 | 28.0157 | 15.0459 | 0.0105 | -1.5165 | 52.1989 |
| 6 | 14.1075 | 21.8879 | 22.4148 | 0.0277 | -3.1275 | 52.2198 |
| 7 | -69.9977 | -45.1863 | 45.7964 | 0.0796 | -2.5036 | 49.2930 |
| 8 | 70.4393 | 122.4190 | 43.1072 | 0.4920 | -2.6342 | 49.6754 |
| 9 | -4.1953 | -2.1201 | 21.2576 | -0.1437 | -1.7988 | 51.6757 |
| 10 | 1.7355 | 2.6144 | 20.7924 | 0.2370 | -2.2927 | 56.3232 |

Note: All data are displayed on a percentage basis, and $r$ = *Naïve average return* + 0.5 × | *Naïve average return*|.

With respect to case 1, the optimal portfolio allocation yields short sales of SP500 index over regimes 3 and 8, while short sales of crude oil occur over regimes 4 and 7. Analogously, the optimal weights of natural gas are negative over regimes 1, 2, 4, 5 and 7. Case 2 further constrains the holdings ($w_l$) in SP500 index to lie between zero and unity. Therefore, under case 2, short sales of SP500 become forbidden. Under such case, the optimal portfolio allocation yields short sales of crude oil over regime 7, while short sales of natural gas occur over regimes 1, 2, 4, 5 and 8. Moreover, the optimal portfolio exhibits high degrees of exposures to commodities over regimes 7 and 8 since some weights lie far above 1 in absolute value. Such features result from the joint dependence structures of portfolio constituents over previous regimes. Strikingly, cases 3 and 4 are alike because the optimal weights of SP500 are always positive and similar in the presence and the absence of short sales. The optimal portfolio attributes negative weights to natural gas over regimes 1, 2 and 5. Finally, when switching from case 1 to case 2, the optimal portfolio's performance often experiences a drop as sketched by the remarked decrease in the upside potential ratio (UPR). Adding a weight constraint to $w_l$ lowers more or less the optimal portfolio's performance.

**Table 5:** Optimal portfolios' attributes under variance minimization

| Regime | Optimization scheme | | | | | | | | | |
|---|---|---|---|---|---|---|---|---|---|---|
| | Case 1 | | | | | Case 2 | | | | |
| | $w_1$ | $w_2$ | $w_3$ | r fixed | UPR | $w_1 \in [0,1]$ | $w_2$ | $w_3$ | r fixed | UPR |
| 1 | 49.0351 | 55.9687 | -5.0038 | -12.7882 | 55.8655 | 49.0351 | 55.9687 | -5.0038 | -12.7882 | 55.8655 |
| 2 | 104.9556 | 9.2022 | -14.1578 | 4.2816 | 53.4644 | 99.9998 | 11.5459 | -11.5457 | 4.2816 | 53.4924 |
| 3 | -7.1090 | 84.7824 | 22.3266 | 54.8640 | 50.4349 | 0.0000 | 69.5692 | 30.4308 | 54.8640 | 49.5003 |
| 4 | 195.2979 | -37.6242 | -57.6737 | 19.8022 | 52.0666 | 99.9999 | 97.2499 | -97.2498 | 19.8022 | 51.5033 |
| 5 | 96.0056 | 12.2824 | -8.288 | 28.0157 | 47.7322 | 96.0049 | 12.2827 | -8.2876 | 28.0157 | 47.7322 |
| 6 | 28.2351 | 57.7848 | 13.9801 | 21.8879 | 57.4723 | 28.2351 | 57.7848 | 13.9801 | 21.8879 | 57.4723 |
| 7 | 144.7155 | -31.6607 | -13.0548 | -45.1863 | 54.9053 | 100.0000 | -479.5437 | 479.5437 | -45.1863 | 54.5021 |
| 8 | -198.1913 | 282.4253 | 15.766 | 122.4190 | 53.8733 | 0.0000 | 887.2497 | -787.2497 | 122.4190 | 47.2812 |
| 9 | 62.9326 | 10.9221 | 26.1453 | -2.1201 | 51.3406 | 62.9326 | 10.9221 | 26.1453 | -2.1201 | 51.3406 |
| 10 | 30.0069 | 45.2647 | 24.7284 | 2.6144 | 54.5746 | 30.0069 | 45.2647 | 24.7284 | 2.6144 | 54.5746 |

| Regime | Case 3 | | | | | Case 4 | | | | |
|---|---|---|---|---|---|---|---|---|---|---|
| | $w_1$ | $w_2$ | $w_3$ | Optimal r | UPR | $w_1 \in [0,1]$ | $w_2$ | $w_3$ | Optimal r | UPR |
| 1 | 94.7498 | 15.3860 | -10.1358 | 21.8820 | 51.0202 | 94.7498 | 15.3860 | -10.1358 | 21.8820 | 51.0202 |
| 2 | 109.0343 | 0.1270 | -9.1613 | 3.2179 | 53.3292 | 100.0000 | 7.0647 | -7.0647 | 3.2179 | 53.2375 |
| 3 | 74.0674 | 24.7614 | 1.1712 | -3.4339 | 56.0803 | 74.0674 | 24.7614 | 1.1712 | -3.4339 | 56.0803 |
| 4 | 86.6125 | 11.4634 | 1.9241 | 15.1230 | 54.9647 | 86.6125 | 11.4634 | 1.9241 | 15.1230 | 54.9647 |
| 5 | 99.0022 | 6.2204 | -5.2226 | 22.6048 | 45.9586 | 99.0022 | 6.2204 | -5.2226 | 22.6048 | 45.9586 |
| 6 | 66.4916 | 25.2662 | 8.2422 | -0.7364 | 54.2025 | 66.4916 | 25.2662 | 8.2422 | -0.7364 | 54.2025 |
| 7 | 36.7157 | 17.0329 | 46.2514 | -69.2184 | 58.0258 | 36.7157 | 17.0329 | 46.2514 | -69.2184 | 58.0258 |
| 8 | 92.5665 | 4.5331 | 2.9004 | 60.7695 | 51.8930 | 92.5665 | 4.5331 | 2.9004 | 60.7695 | 51.8930 |
| 9 | 69.1390 | 12.7327 | 18.1283 | 0.4906 | 49.6325 | 69.1390 | 12.7327 | 18.1283 | 0.4906 | 49.6325 |
| 10 | 81.8862 | 7.9604 | 10.1534 | 10.1318 | 51.6016 | 81.8862 | 7.9604 | 10.1534 | 10.1318 | 51.6016 |

Note: All data are displayed on a percentage basis. The target return *r* corresponds to the average annual return of the optimal portfolio during each regime.

*5.2. Downside risk optimization*

Table 6 displays the results of semi-variance optimization while Table 7 introduces the results of tail risk optimization.

*Semi-variance optimization results*

With respect to case 1, the optimal portfolio allocation yields short sales of SP500 index over regimes 3 and 8, while short sales of crude oil occur over regimes 4 and 7. Analogously, the optimal weights of natural gas are negative over regimes 2, 4, 5 and 7. With respect to case 2, the optimal portfolio allocation yields short sales of crude oil over regime 7, while short sales of natural gas occur over regimes 2, 4, 5 and 8. Amazingly, cases 3 and 4 are analogous since the optimal weights of SP500 are always positive and similar in the presence and the absence of short sales. The optimal portfolio sells short natural gas over regimes 1, 2, 3 and 5. Finally, when switching from case 1 to case 2, the optimal portfolio's performance often experiences a drop since the upside potential ratio (UPR) decreases. Adding a weight constraint to $w_1$ weakens more or less the optimal portfolio's performance.

**Table 6:** Optimal portfolios' attributes under semi-variance minimization

| Regime | Optimization scheme | | | | | | | | | |
|---|---|---|---|---|---|---|---|---|---|---|
| | Case 1 | | | | | Case 2 | | | | |
| | $w_1$ | $w_2$ | $w_3$ | *r fixed* | UPR | $w_1 \in [0,1]$ | $w_2$ | $w_3$ | *r fixed* | UPR |
| 1 | 49.7258 | 46.8754 | 3.3988 | -12.7882 | 54.6875 | 49.7259 | 46.8743 | 3.3998 | -12.7882 | 54.6875 |
| 2 | 96.0216 | 13.4273 | -9.4489 | 4.2816 | 53.4529 | 96.0216 | 13.4273 | -9.4489 | 4.2816 | 53.4529 |
| 3 | -10.1808 | 91.3562 | 18.8246 | 54.8640 | 50.6329 | 0.0000 | 69.5692 | 30.4308 | 54.8640 | 49.5003 |
| 4 | 197.4440 | -40.6615 | -56.7825 | 19.8022 | 52.2648 | 100.0000 | 97.2499 | -97.2499 | 19.8022 | 51.5033 |
| 5 | 86.8828 | 16.6318 | -3.5146 | 28.0157 | 47.1954 | 86.8833 | 16.6316 | -3.5149 | 28.0157 | 47.1954 |
| 6 | 28.1865 | 57.6903 | 14.1232 | 21.8879 | 57.4673 | 28.1865 | 57.6903 | 14.1232 | 21.8879 | 57.4673 |
| 7 | 144.6432 | -32.3855 | -12.2577 | -45.1863 | 54.9417 | 100.0000 | -479.5437 | 479.5437 | -45.1863 | 54.5021 |
| 8 | -197.7362 | 283.8142 | 13.922 | 122.4190 | 53.7706 | 0.0000 | 887.2497 | -787.2497 | 122.4190 | 47.2812 |
| 9 | 62.0875 | 11.7443 | 26.1682 | -2.1201 | 51.3572 | 62.0875 | 11.7443 | 26.1682 | -2.1201 | 51.3572 |
| 10 | 29.8267 | 45.5726 | 24.6007 | 2.6144 | 54.5453 | 29.8269 | 45.5723 | 24.6008 | 2.6144 | 54.5453 |

| Regime | Case 3 | | | | | Case 4 | | | | |
|---|---|---|---|---|---|---|---|---|---|---|
| | $w_1$ | $w_2$ | $w_3$ | *Optimal r* | UPR | $w_1 \in [0,1]$ | $w_2$ | $w_3$ | *Optimal r* | UPR |
| 1 | 85.2434 | 17.0629 | -2.3063 | 13.6597 | 49.6661 | 85.2547 | 17.0533 | -2.308 | 13.2431 | 49.8123 |
| 2 | 98.9309 | 5.7625 | -4.6934 | 3.3449 | 53.0777 | 98.9308 | 5.7626 | -4.6934 | 3.3449 | 53.0776 |
| 3 | 75.8662 | 24.2763 | -0.1425 | -4.7753 | 56.0745 | 75.8662 | 24.2763 | -0.1425 | -4.7753 | 56.0745 |
| 4 | 86.0354 | 11.9690 | 1.9956 | 15.1094 | 55.0427 | 86.0354 | 11.9698 | 1.9948 | 15.1094 | 55.0426 |
| 5 | 91.0624 | 10.1704 | -1.2328 | 22.7929 | 45.2130 | 91.0558 | 10.1736 | -1.2294 | 22.7932 | 45.2126 |
| 6 | 65.9386 | 25.3624 | 8.699 | -0.5128 | 54.2495 | 65.9386 | 25.3624 | 8.699 | -0.5128 | 54.2495 |
| 7 | 36.9546 | 16.8717 | 46.1737 | -69.1782 | 58.0151 | 36.9402 | 16.8783 | 46.1815 | -69.1805 | 58.0157 |
| 8 | 91.0242 | 6.2251 | 2.7507 | 61.0565 | 51.8674 | 90.9988 | 6.2495 | 2.7517 | 61.0610 | 51.8680 |
| 9 | 69.1055 | 12.7790 | 18.1155 | 0.4951 | 49.6313 | 69.1058 | 12.7788 | 18.1154 | 0.4951 | 49.6313 |
| 10 | 79.3453 | 10.3521 | 10.3026 | 9.8365 | 51.7470 | 79.3455 | 10.3520 | 10.3025 | 9.8365 | 51.7470 |

Note: All data are displayed on a percentage basis. The target return *r* corresponds to the average annual return of the optimal portfolio during each regime.

*Tail risk optimization results*

With respect to case 1, the optimal portfolio allocation yields short sales of SP500 index over regimes 3 and 8, while short sales of crude oil occur over regimes 4 and 7. Analogously, the optimal weights of natural gas are negative over regimes 2, 4, 5, 7 and 8. With respect to case 2, the optimal portfolio allocation yields short sales of crude oil over regime 7, while short sales of natural gas occur over regimes 2, 4, 5 and 8. Moreover, the optimal portfolio exhibits high degrees of exposures to commodities over regimes 7 and 8 since some weights lie far above 1 in absolute value. Such features result from the joint dependence structures of portfolio constituents over previous regimes. The three constituent assets follow a Gaussian copula, which highlights the absence of tail dependence. Moreover, energy commodities are positively correlated with SP500 index over these regimes. Surprisingly, cases 3 and 4 are comparable because the optimal weights of SP500 are always positive and similar in the absence and the presence of short sales. The optimal weights of natural gas are negative over regimes 1, 2, 3 and 5. Finally, when switching from case 1 to case 2, the optimal portfolio's performance repeatedly experiences a drop as suggested by the remarked decrease in the upside potential ratio (UPR). Adding a weight constraint to $w_1$ moderately lessens the optimal portfolio's performance.

*Comparing weights' signs across risk measures and optimization constraints*

When switching from case 1 to case 2 (i.e. under a return constraint coupled first with, and then, without SP500 short sales), the signs of optimal weights mainly persist under all risk measures. Many negative signs disappear when switching from case 1 to case 3, and from case 2 to case 4 (i.e. when relaxing the constraint on the expected return of the optimal portfolio). Negative weights prevail only for natural gas over regimes 1, 2, 3 (sometimes), and 5. Therefore, without a constraint on its expected return, the optimal portfolio allows less short sales. Under the variance measure and without a return constraint, both the low variance regime of natural gas, and the Gumbel dependence structure (i.e. positive correlation, and upper tail dependence) drive the persisting negative weights. Similarly, under the semi-variance and tail risk measures, the asymmetric tail dependence (i.e. Gumbel and Clayton copulas, namely upper and lower tail dependence, with positive correlation) drives negative weights.

**Table 7:** Optimal portfolios' attributes under tail risk minimization

| Regime | Optimization scheme | | | | | | | | | |
|---|---|---|---|---|---|---|---|---|---|---|
| | Case 1 | | | | | Case 2 | | | | |
| | $w_1$ | $w_2$ | $w_3$ | r fixed | UPR | $w_1 \in [0,1]$ | $w_2$ | $w_3$ | r fixed | UPR |
| 1 | 49.7132 | 47.0408 | 3.2460 | -12.7882 | 54.6908 | 49.7132 | 47.0408 | 3.2460 | -12.7882 | 54.6908 |
| 2 | 97.5097 | 12.7236 | -10.2333 | 4.2816 | 53.4841 | 97.4934 | 12.7313 | -10.2247 | 4.2816 | 53.4838 |
| 3 | -11.5000 | 94.1793 | 17.3207 | 54.8640 | 50.6862 | 13.6264 | 40.4086 | 45.9650 | 54.8640 | 45.9131 |
| 4 | 175.6805 | -9.8599 | -65.8206 | 19.8022 | 50.5295 | 100.0000 | 97.2498 | -97.2498 | 19.8022 | 51.5033 |
| 5 | 85.0990 | 17.4823 | -2.5813 | 28.0157 | 47.3105 | 85.0997 | 17.4819 | -2.5816 | 28.0157 | 47.3104 |
| 6 | 27.7572 | 56.8549 | 15.3879 | 21.8879 | 57.4101 | 27.7572 | 56.8548 | 15.3880 | 21.8879 | 57.4101 |
| 7 | 144.8260 | -30.5539 | -14.2721 | -45.1863 | 54.8651 | 100.0000 | -479.5436 | 479.5436 | -45.1863 | 54.5021 |
| 8 | -193.8955 | 295.5349 | -1.6394 | 122.4190 | 53.2764 | 0.0000 | 887.2497 | -787.2497 | 122.4190 | 47.2812 |
| 9 | 63.8463 | 10.0330 | 26.1207 | -2.1201 | 51.3370 | 63.8373 | 10.0417 | 26.1210 | -2.1201 | 51.3370 |
| 10 | 28.1006 | 48.5223 | 23.3771 | 2.6144 | 54.2337 | 28.0996 | 48.5240 | 23.3764 | 2.6144 | 54.2335 |
| Regime | Case 3 | | | | | Case 4 | | | | |
| | $w_1$ | $w_2$ | $w_3$ | Optimal r | UPR | $w_1 \in [0,1]$ | $w_2$ | $w_3$ | Optimal r | UPR |
| 1 | 94.4447 | 9.7504 | -4.1951 | 21.1828 | 49.8614 | 94.4448 | 9.7503 | -4.1951 | 21.1828 | 49.8614 |
| 2 | 100.3605 | 5.0938 | -5.4543 | 3.3459 | 53.1214 | 100.0000 | 5.3306 | -5.3306 | 3.3558 | 53.1154 |
| 3 | 74.2472 | 25.8197 | -0.0669 | -4.0114 | 55.9909 | 74.2477 | 25.8191 | -0.0668 | -4.0116 | 55.9910 |
| 4 | 85.9032 | 12.1499 | 1.9469 | 15.1091 | 55.0486 | 85.9032 | 12.1499 | 1.9469 | 15.1091 | 55.0486 |
| 5 | 85.0573 | 17.5863 | -2.6436 | 28.1159 | 47.3591 | 85.0613 | 17.5842 | -2.6455 | 28.1156 | 47.3587 |
| 6 | 60.7309 | 29.6918 | 9.5773 | 2.2889 | 54.8106 | 60.7309 | 29.6917 | 9.5774 | 2.2889 | 54.8106 |
| 7 | 32.1892 | 18.8225 | 48.9883 | -69.9509 | 58.0656 | 32.1892 | 18.8225 | 48.9883 | -69.9509 | 58.0656 |
| 8 | 89.4893 | 7.6326 | 2.8781 | 61.3302 | 51.9352 | 89.4893 | 7.6326 | 2.8781 | 61.3302 | 51.9352 |
| 9 | 68.7169 | 12.8808 | 18.4023 | 0.4022 | 49.6934 | 68.7177 | 12.8803 | 18.402 | 0.4022 | 49.6934 |
| 10 | 81.6750 | 7.8548 | 10.4702 | 10.0611 | 51.6827 | 81.6753 | 7.8545 | 10.4702 | 10.0611 | 51.6827 |

Note: All data are displayed on a percentage basis. The target return $r$ corresponds to the average annual return of the optimal portfolio during each regime.

*5.3. Comparing optimal portfolios and performance diagnostics*

We first compare the compositions of optimal portfolios through a well-chosen distance measure. Then, we undertake a performance diagnostic based on cumulative returns and expected maximum drawdowns. The cumulative return reflects the performance of optimal portfolios, whereas the expected maximum drawdown represents a measure for optimal portfolios' risk.

The potential heterogeneity, or equivalently, dissimilarity of obtained optimal portfolios is assessed through a well-chosen distance measure, across the 10 referenced periods of time. Analogously to Fulga (2016),[5] we consider the $L_1$-norm as a distance measure. We define the average distance measure as the following dissimilarity index (DI):

$$DI(j,k) = \frac{1}{n}\sqrt{\sum_{i=1}^{n}\left(w_i^j - w_i^k\right)^2} \qquad (9)$$

where the number of constituent assets $n$ is equal to 3, and $j$ and $k$ represent a pair of cases among the four possible cases (i.e. four different optimal portfolios). Setting $j$ equal to $k$ is equivalent to calculate the dissimilarity index of the same portfolio. In such situation, we have $w_i^j - w_i^k = 0$ whatever the considered asset $i$. As a result, the dissimilarity index is zero. Besides, the larger the dissimilarity index is, the more heterogeneous and dissimilar portfolios become. Since cases 3 and 4 are very close, we assimilate case 4 to case 3, so that finally we consider only 3 cases. As reported in Table 8, optimal portfolios are generally heterogeneous under all risk measures. However, the optimal portfolios of cases 1 and 2 become homogenous over regimes 1, 2, 5, 6, 9 and 10. Such findings coincide with previous results because case 2 offsets only the negative SP500 weights of case 1 over regimes 3 and 8. Over both periods, the variance regimes of SP500, crude oil and natural gas are similar (i.e. M/M/H). The high variance regime of natural gas prices seems to drive the negative SP500 weights. Finally, under cases 1 and 3/4, the tail-risk optimal portfolios are very close over period 5.

**Table 8:** Dissimilarity index across periods and risk measures

| | Pairs *j-k* | 1 | 2 | 3 | 4 | 5 | 6 | 7 | 8 | 9 | 10 |
|---|---|---|---|---|---|---|---|---|---|---|---|
| Variance | 1-2 | 0.0000 | 2.0242 | 6.2152 | 56.6068 | 0.0003 | 0.0000 | 222.4240 | 341.5531 | 0.0000 | 0.0000 |
| Variance | 1-3/4* | 20.4481 | 3.7112 | 34.3829 | 44.4397 | 2.4749 | 16.8455 | 44.1616 | 134.1350 | 3.4330 | 21.8467 |
| Variance | 2-3/4* | 14.6316 | 33.0336 | 18.0594 | 36.8953 | 31.3325 | 6.5777 | 233.0476 | 376.0505 | 21.0267 | 13.6591 |
| Semi-variance | 1-2 | 0.0005 | 0.0000 | 8.9008 | 57.8816 | 0.0002 | 0.0000 | 222.0641 | 340.7688 | 0.0000 | 0.0001 |
| Semi-variance | 1-3/4* | 15.5736 | 3.1593 | 36.9136 | 45.5054 | 2.6755 | 16.6658 | 44.0168 | 133.5677 | 3.5772 | 20.8087 |
| Semi-variance | 2-3/4* | 14.0188 | 29.1029 | 18.5689 | 37.0759 | 27.0446 | 6.3971 | 233.1386 | 376.8138 | 20.7607 | 12.4844 |

---

[5] The author defines a dissimilarity index for a two-asset portfolio whose weights lie only between 0 and 1. Hence, the dissimilarity index is bounded between 0 and 1.

|         | Pairs j-k | 1       | 2       | 3       | 4       | 5       | 6       | 7        | 8        | 9       | 10      |
|---------|-----------|---------|---------|---------|---------|---------|---------|----------|----------|---------|---------|
| Tail risk | 1-2     | 0.0000  | 0.0067  | 21.9675 | 44.9541 | 0.0003  | 0.0000  | 222.9736 | 334.1499 | 0.0042  | 0.0007  |
|         | 1-3/4*    | 19.5700 | 3.1478  | 37.0104 | 38.2053 | 0.0427  | 14.3715 | 46.1001  | 134.6663 | 3.1869  | 22.8294 |
|         | 2-3/4*    | 16.2045 | 29.8297 | 22.5297 | 37.1421 | 25.6324 | 6.4220  | 231.6614 | 377.4990 | 21.1850 | 12.4794 |

* Cases 3 and 4 are similar so that they reduce to one case.

As regards cumulative returns, Table 9 displays the best-performing optimal portfolios across risk measures, over each listed period, whereas Table 10 displays annualized cumulative returns across cases. Table 10 presents the cumulative returns over the whole sample period, under each case. Table 9 displays the best-performing portfolio, which presents the highest annualized cumulative return across the three risk measures, over a given period. The optimal portfolios offering the highest period-specific annualized cumulative returns result from variance, semi-variance and tail risk minimization in 25%, 52.50% and 35% of situations respectively. The optimal portfolios resulting from tail risk minimization generally outperform, when SP500, crude oil and natural gas present low variance regimes. The optimal portfolios resulting from semi-variance minimization offer the best annualized cumulative returns, when SP500, crude oil and natural gas display diverse and higher variance regimes. In the remainder of situations, variance optimal portfolios outperform. Additionally, the cumulative returns in Table 10 reflect the performance of a period-specific portfolio rebalancing strategy, across the four possible cases. Under any given case, portfolio optimization/rebalancing occurs over each regime within the sample period, and the same optimization criterion applies to each regime. Such regime-specific active portfolio management captures the portfolio's structural changes over time, namely the time-varying dependence structure of its assets. Tail-risk optimal portfolios outperform under cases 3 and 4, when the expected return is unconstrained. Conversely, the optimal portfolios resulting from the semi-variance minimization outperform under case 1, when the expected return is fixed (i.e. when a performance target is imposed). Finally, the positive weight constraint on SP500 index explains the poor performance of optimal portfolios under case 2. Banning short sales prevents the hedging role of energy commodities, specifically over periods 7 and 8 (see Figure 2). During these periods, the SP500 index presents a positive correlation with energy commodities. Such correlation strongly alters the optimal portfolio's performance.

**Table 9:** Best-performing portfolios with highest annualized cumulative returns

| Period | 1  | 2  | 3  | 4 |
|--------|----|----|----|---|
| 1      | SV | SV | V  | V |
| 2      | SV | SV | SV | V |
| 3      | SV | SV | V  | V |

| Period | 1 | 2 | 3 | 4 |
|---|---|---|---|---|
| 4 | SV | V-SV | SV | SV |
| 5 | TR | TR | TR | TR |
| 6 | TR | TR | TR | TR |
| 7 | SV | V-SV-TR | SV | SV |
| 8 | SV | V-SV-TR | TR | TR |
| 9 | TR | TR | SV | SV |
| 10 | SV | SV | V | V |

Note: V, SV and TR stand for variance, semi-variance and tail risk. When more than one risk measure appears, all mentioned risk measures yield the same result.

**Table 10:** Annualized cumulative returns of optimal portfolios across cases (%)

| Case/Measure | 1 | 2 | 3 | 4 |
|---|---|---|---|---|
| Variance | 3.2493 | -4.2974 | 2.3243 | 2.5313 |
| Semi-variance | **3.4628** | -4.2974 | 1.9094 | 1.9098 |
| Tail Risk | 3.1669 | -4.2974 | **2.6384** | **2.6451** |

Note: Highest cumulative returns are in bold.

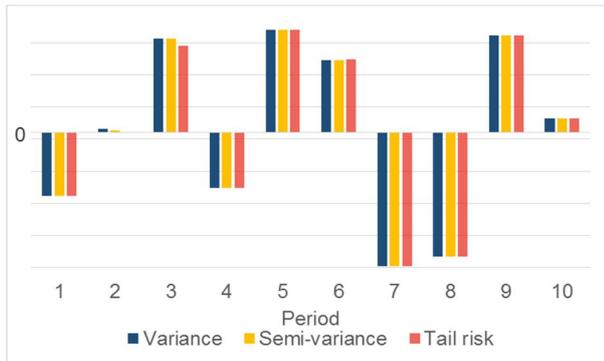

**Figure 2:** Cumulative returns of optimal portfolios under a positive SP500 weight constraint

As regards drawdowns, the maximum drawdown represents the maximum loss a portfolio can experience in value or percentage (i.e. percentage drawdown), over a given period of time (e.g. investment horizon). It is measured as the difference between a peak and a nadir value of the portfolio. The maximum drawdown depends on the portfolio's average return and returns' fluctuations. The expected maximum drawdown represents the expected value of the maximum drawdown. Specifically, the expected maximum drawdown is computed while assuming the portfolio's cumulative returns to follow a Brownian motion (e.g. a random walk). Under such assumption, cumulative returns exhibit a constant drift over time with random deviations from such drift (Magdon-Ismail et al., 2004). The expected maximum drawdown depends positively on both the investment horizon and return volatility, but negatively on the portfolio's expected return. Table 11 displays the expected maximum drawdowns of optimal portfolios over the whole sample, and under the four considered cases. The expected maximum drawdowns reflect the risk of optimally rebalanced portfolios under the considered cases. Under cases 1 and 2, semi-variance optimal portfolios exhibit the lowest expected maximum drawdowns, when the expected return of the optimal portfolio is constrained. Differently, tail-risk optimal portfolios exhibit the lowest cumulative losses under cases 3 and 4, when the expected return of the optimal portfolio is unconstrained. The strongest loss scenario happens under case 2, when both the expected return of the optimal portfolio is fixed, and SP500 weight forbids short sales. The weakest loss scenario happens under cases 3 and 4. Thus, the constraint on the expected return reinforces the expected maximum drawdown. Such effect is magnified when we add a positivity constraint to SP500 weight. When risk is measured by the expected maximum drawdown, and performance is measured by the cumulative return, the tail-risk optimal portfolio offers therefore the best risk-return tradeoff under cases 3 and 4. Analogously, the semi-variance optimal portfolio offers the best risk-return tradeoff under case 1 (see Figure 3).

**Table 11:** Expected maximum drawdowns of optimal portfolios across cases (%)

| Case/Measure | 1 | 2 | 3 | 4 |
|---|---|---|---|---|
| Variance | 187.3206 | 406.1854 | 82.2172 | 78.5013 |
| Semi-variance | **184.2593** | **406.0575** | 79.7267 | 79.7259 |
| Tail Risk | 188.3034 | 406.9348 | **77.7832** | **77.6408** |

Note: Lowest expected maximum drawdowns are in bold.

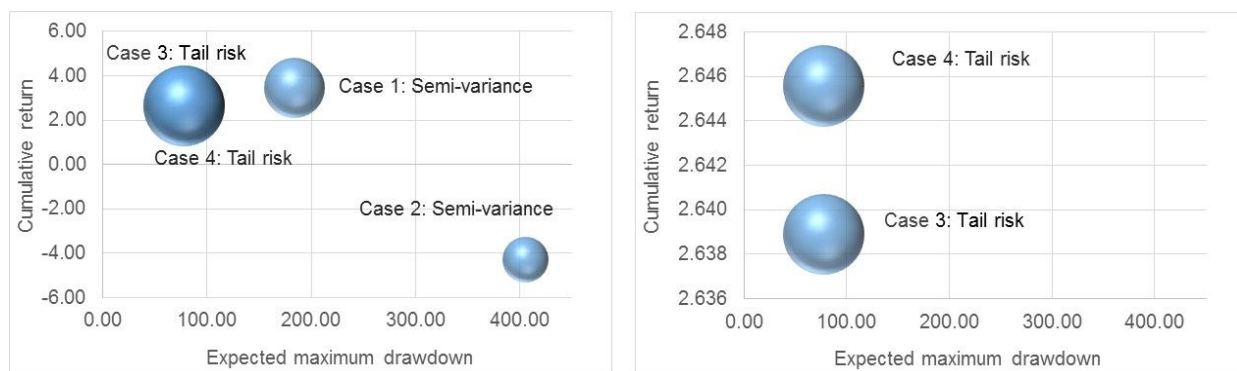

**Figure 3:** Risk-return tradeoff of optimal portfolios over the whole sample period

Note: Data are expressed in percent. The size of the bubbles represents the ratio of the absolute value of the cumulative return to the expected maximum drawdown.

For comparison purposes, we also carry out the active management strategy, which results from Table 9. Over each regime, we build/rebalance the portfolio according to the most favorable risk minimization setting (i.e. variance, semi-variance or tail-risk optimal weights). Such methodology allows for building superoptimal portfolios over the whole sample period. Such portfolios are optimally rebalanced across regimes and risk measures. Table 12 displays the risk and return attributes of the superoptimal portfolios. The annualized cumulative returns and expected maximum drawdowns of these superoptimal portfolios are close to the ones in Table 10 and Table 11 (see Figure 4). Due to the difficulty to forecast dependence structures and suitable risk minimization, investors can rely on previous findings since optimal and superoptimal portfolios offer close risk-return tradeoffs. As a result, active portfolio managers should build semi-variance optimal portfolios in the presence of a return constraint. Without such constraint, they should build tail-risk optimal portfolios.

**Table 12:** Attributes of superoptimal portfolios across cases (%)

| Case | 1 | 2 | 3 | 4 |
|---|---|---|---|---|
| Cumulative Return | 3.4639 | -4.2974 | 2.8771 | 2.9120 |
| Expected maximum drawdown | 184.2438 | 406.0562 | 76.2858 | 76.8912 |

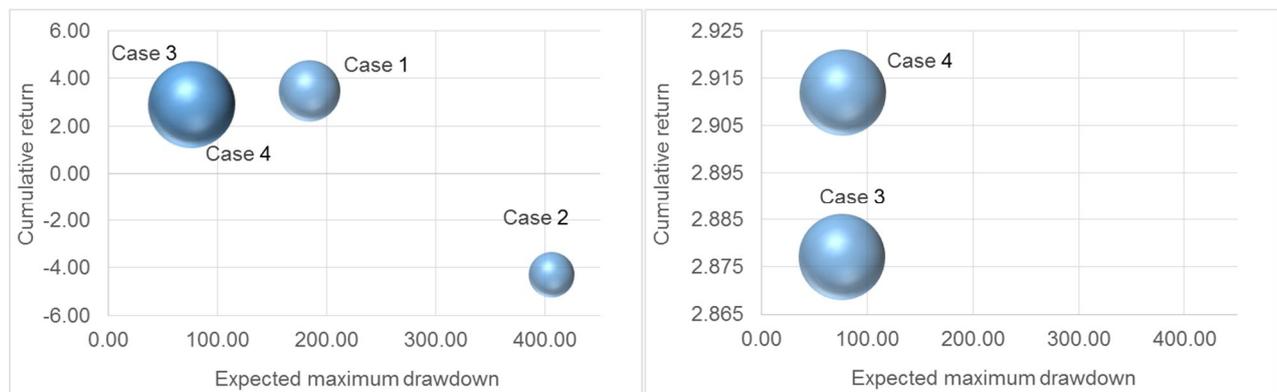

**Figure 4:** Risk-return tradeoff of superoptimal portfolios over the whole sample period

Note: Data are expressed in percent. The size of the bubbles represents as the ratio of the absolute value of the cumulative return to the expected maximum drawdown.

## 6. Conclusion

We consider the joint behavior of energy commodity prices and the U.S. stock market index over time. The SP500 index as well as crude oil and natural gas prices exhibit structural changes with various variance regimes. As a result, the joint dependence structure of the U.S. stock market index and the two energy commodities is unstable over time. Such feature has significant implications for investors building portfolios with these three types of assets, and requires a regime-specific analysis. In this light, we examine the implications of such regime-dependency for portfolio optimization. We minimize various risk measures after accounting for regime-specific dependence structures. The dependence structures handle the asymmetry in asset returns and tail dependency. The risk criteria consist of variance, and two downside risk measures such as semi-variance and tail risk. Moreover, we consider several minimization schemes while setting or discarding constraints on the portfolio's expected return and SP500 weight. We examine optimal portfolios while analyzing their degrees of similarity, their cumulative returns as a performance indicator, and their expected maximum drawdowns as a risk measure.

The positive weight constraint on SP500 index moderately reduces the performance of the optimal portfolio over given regimes. Besides, optimal portfolios are heterogeneous and dissimilar across regimes and risk minimization schemes. We also build optimal portfolios, which we rebalance across regimes, and analyze their risk-return tradeoff over the whole sample period. Under each risk minimization scheme, the expected maximum drawdown serves as a risk measure while the cumulative return serves as a performance measure. Under a constraint on the expected return of the optimal portfolio, the semi-variance optimal portfolio offers the best risk-return tradeoff (i.e. lowest risk and highest return). Differently, the tail-risk optimal portfolio offers the best risk-return tradeoff without constraint on the expected return. As a consequence, optimization results handle regime-dependency since they depict the joint risk structure of portfolio constituents over time. They capture the price uncertainty of both constituent assets and resulting portfolio over time, and highlight the efficiency of an active portfolio management strategy. Future research should exploit such findings, and attempt to forecast the upcoming portfolio scenarios. Potential research extensions rely on a scenario analysis, which faces two major challenges. First, the analysis needs to predict the upcoming variance regimes of constituent assets, and then those of the portfolio. Second, such approach needs to identify the plausible dependence structures over the forthcoming variance regimes. For example, such scenario analysis can exploit the plurality of past variance regimes and corresponding dependence structures.

**Acknowledgements** This work was partly achieved through the Laboratory of Excellence on Financial Regulation (Labex ReFi) supported by PRES heSam under the reference ANR10LABX0095. It benefited from a French government support managed by the National Research Agency (ANR) within the project Investissements d'Avenir Paris Nouveaux Mondes (investments for the future Paris New Worlds) under the reference ANR11IDEX000602.

**References:**

Alexander, C. (1998): Volatility and Correlation: Measurement, Models and Applications, in: Risk Management and Analysis: Measuring and Modelling Financial Risk (Volume 1), C. Alexander (eds), John Wiley & Sons: England, Chapter 4, 125-172.

Aloui, R., Safouane Ben Aïssa, M., Hammoudeh, S. and Nguyen, D. K. (2014): Dependence and extreme dependence of crude oil and natural gas prices with applications to risk management, Energy Economics, 42, 332–342.

Barberis, N. C., and Huang, M. (2008): Stocks as lotteries: The implications of probability weighting for security prices, American Economic Review, 98(5), 2066–2100.

Barsky, R. B. and Kilian L. (2004): Oil and the Macroeconomy Since the 1970s, Journal of Economic Perspectives, 18(4), 115-134.

Bekkers, N., Doeswijk, R. Q. and Lam, T. W. (2009): Strategic asset allocation: determining the optimal portfolio with ten asset classes, Journal of Wealth Management, 12 (3), 61-77.

Brigida, M. (2014): The switching relationship between natural gas and crude oil prices, Energy Economics, 43 (C), 48-55.

Brown, S.P.A. and Yücel, M.K. (2008): What drives natural gas prices?, Energy Journal, 29 (2), 45-60.

Buckley, I., Saunders, D. and Seco, L. (2008): Portfolio optimization when asset returns have the Gaussian mixture distribution. European Journal of Operational Research, 185, 1434-1461.

Büyükşahin, B., Haigh, M. S. and Robe, M. A. (2010): Commodities and equities: Ever a "Market of One"?, Journal of Alternative Investments, 12(3), 76–95.

Cherubini, U., Luciano, E. and Vecchiato, W. (2004 ): Copula Methods in Finance, Chichester: Wiley.

Cheung, C. S. and Miu, P. (2010): Diversification benefits of commodity futures, Journal of International Financial Markets, Institutions, and Money, 20 (5), 451-474.

Chevallier, J. and Ielpo, F. (2013): Economic regimes and commodity markets as an asset class, in: The Economics of Commodity Markets, Wiley, U.K., Chapter 4, 145-178.

Chevallier, J., Gatumel, M. and Ielpo, F. (2014): Commodity markets through the business cycle, Quantitative Finance, 14 (9), 1597-1618.

Christopherson, R., Gregoriou, G. N. and Struga, A. (2004): Optimal allocation of commodity trading advisors in an international stock, bond and hedge fund portfolio. Derivatives Use, Trading & Regulation, 10 (3), 229-239.

Conlon, T. and Cotter, J. (2013): Downside Risk and the Energy Hedger's Horizon, Energy Economics, 36 (March), 371–379.

Conover, C. M., Jensen, G. R., Johnson, R. R. and Mercer, J. M. (2010): Is now the time to add commodities to your portfolio? Journal of Investing, 19 (3), 10-19.

Daigler, R. T., Dupoyet, B. and You, L. (2016): Spicing Up a Portfolio with Commodity Futures: Still a Good Recipe? Forthcoming in the Journal of Alternative Investments.

Davidon, W. C. (1991): Variable Metric Method for Minimization, SIAM Journal on Optimization, 1 (1), 1–17.

Deheuvels, P. (1979): La Fonction de Dépendance Empirique et ses Propriétés, un Test Non Paramétrique d'indépendance. Bulletin de l'Académie Royale de Belgique, Classe des Sciences, 65 (6), 274-292.

Domanski, D. and Heath, A. (2007): Financial investors and commodity markets. BIS Quarterly Review, 53–67.

Dowd, K. and Blake, D. (2006): After VaR: The Theory, Estimation, and Insurance Applications of Quantile-Based Risk Measures, Journal of Risk and Insurance, 73 (2), 193-229.

Eckaus, R.S. (2008): The Oil Price Really is a Speculative Bubble, Working Paper 08-007WP, Massachusetts Institute of Technology, Center for Energy and Environmental Policy Research.

Embrechts, P., Lindskog, F. and McNeil, A.J. (2003 ): Modeling dependence with copulas and applications to risk management, In: Handbook of Heavy Tailed Distributions in Finance, S. Rachev, pp. 329–384, Elsevier, North-Holland.

Falkowski, M. (2011). Financialization of commodities. Contemporary Economics, 5 (4), 4-17.

Fletcher, R. and Powell, M. J. D . (1963): A Rapidly Convergent Descent Method for Minimization, Computer Journal, 6 (2), 163-168.

Gatfaoui, H. (2010): Investigating the dependence structure between credit default swap spreads and the U.S. financial market, Annals of Finance, 6(4), 511-535.

Gatfaoui, H. (2015): Linking the gas and oil markets with the stock market: Investigating the U.S. relationship, Forthcoming in Energy Economics, doi:10.1016/j.eneco.2015.05.021.

Gatfaoui, H. (2016): Capturing Long-Term Coupling and Short-Term Decoupling Crude Oil and Natural Gas Prices, ECOMFIN2016: Energy & Commodity Finance Conference 2016, ESSEC Business School, Paris, June 23-24, 2016.

Georgiev, G. (2001): Benefits of commodity investment Journal of Alternative Investments, 4 (1), 40–48.

Gorton, G. and Rouwenhorst, K. G. (2006): Facts and Fantasies about Commodity Futures, Financial Analysts Journal, 62 (2), 47-68.

Greer, R. J. (2006): Commodity indexes for real return, in: R. J. Greer (Eds), The Handbookof Inflation Hedging Investments (McGraw Hill, U.S.A.), Chapter 5, 105-126.

Hamilton, J. D. (1985): Historical Causes of Postwar Oil Shocks and Recessions, The Energy Journal, 6(1), 97-116.

Hartley, P. R. and Medlock, K. B. (2014): The relationship between crude oil and natural gas prices: The role of the exchange rate, Energy Journal, 35 (2), 25-44.

Harvey, C. R. and Erb, C. B. (2006): The strategic and tactical value of commodity futures, Financial Analysts Journal, 62 (2), 69-97.

Harvey, C. R. and Siddique, A. (2000): Conditional skewness in asset pricing tests, Journal of Finance, 55 (3), 1263–1295.

Hensel, C. R. and Ankrim, E. M. (1993): Commodities in asset allocation: A real-asset alternative to real estate?, Financial Analysts Journal, 49 (3), 20-29.

Jarrow, R. and Zhao, F. (2006): Downside Loss Aversion and Portfolio Management, Management Science, 52 (4), 558-566.

Jensen, G. R., Johnson, R. R. and Mercer, J. M. (2000): Efficient use of commodity futures in diversified portfolios, Journal of Futures Markets, 20 (5), 489-506.

Jin, Y. and Jorion, P. (2006): Firm Value and Hedging: Evidence from U.S. Oil and Gas Producers, Journal of Finance, 61 (2), 893–919.

Kakouris, I. and Rustem, B. (2014): Robust Portfolio Optimization with Copulas, European Journal of Operational Research, 235 (1), 28-37.

Kaplan, P. D. and Lummer, S. L. (1997): GSCI collateralised futures as a hedging and diversification tool for institutional investors: An update, Journal of Investing, 7 (4), 11-17.

Karolyi, G. A., Lee K.-H. and van Dijk, M. (2012): Understanding Commonality in Liquidity Around the World, Journal of Financial Economics, 105 (1), 82-112.

Kilian, L. (2009): Not All Oil Price Shocks are Alike: Disentangling Demand and Supply Shocks in the Crude Oil Market, American Economic Review, 99 (3), 1053-1069.

Kilian, L. and Park, C. (2009): The Impact of Oil Price Shocks on the U.S. Stock Market, International Economic Review, 50(4), 1267-1287.

Kim, M. J., Kim, S., Jo, Y. H. and Kim, S. Y. (2011): Dependence Structure of the Commodity and Stock Markets, and Relevant Multi-Spread Strategy, Physica A, 390 (21-22), 3842-3854.

Kolm, P. N., Tütüncü, R. and Fabozzi, F. J. (2014): 60 years of portfolio optimization: Practical challenges and current trends. European Journal of Operational Research, 234, 356-371.

Kraus, A. and Litzenberger, R. H. (1976): Skewness preference and the valuation of risk assets, Journal of Finance, 31(4), 1085–1100.

Lari-Lavassani, A. and Li, X. (2003): Dynamic Mean Semi-variance Portfolio Selection, Lecture Notes in Computer Science Volume 2657, 95-104.

Lee, C. F. and Leuthold, R. M. (1985): The stock market and the commodity futures market: Diversification and arbitrage potential, Financial Analysts Journal, 41 (4), 53-60.

Lintner, J. (1965): The Valuation of Risk Assets and the Selection of Risky Investments in Stock Portfolios and Capital Budgets, Review of Economics and Statistics, 47 (1), 13-37.

Magdon-Ismail, M., Atiya, A. F., Pratap, A. and Abu-Mostafa, Y. S. (2004): On the Maximum Drawdown of a Brownian Motion, Journal of Applied Probability, 41 (1), 147-161.

Mansini, R., Ogryczak, W. and Speranza, M. G. (2014): Twenty years of linear programming based portfolio optimization. European Journal of Operational Research, 234, 518-535.

Markowitz, H.M. (1952): Portfolio Selection, Journal of Finance, 7 (1), 77–91.

Markowitz, H.M. (1959): Portfolio Selection: Efficient Diversification of Investments, John Wiley & Sons, New York.

McNeil, A., Frey, R. and Embrechts, P. (2005): Quantitative Risk Management: Concepts, Techniques and Tools, Princeton University Press, Princeton.

Miller, J. I. and Ratti, R. A. (2009): Crude Oil and Stock Markets: Stability, Instability, and Bubbles, Energy Economics 31 (4), 559-568.

Mitton, T., and Vorkink, K. (2007): Equilibrium underdiversification and the preference for skewness, Review of Financial Studies, 20 (4), 1255–1288.

Parsons, J.E. (2010): Black Gold and Fool's Gold: Speculation in the Oil Futures Market, Economia 10 (2), 81-116.

Patton, A. W. (2009): Copula–based models for financial time series, In: Handbook of Financial Time Series, Part 5, T. Mikosch, J.-P. Kreiß, R. A. Davis and T. G. Andersen, pp. 767-785, Springer, Berlin Heidelberg.

Pindyck, R. S. (2004): Volatility in Natural Gas and Oil Markets, Journal of Energy and Development, 30 (1), 1-19.

Ramberg, D. J. and Parsons, J. E. (2012): The weah tie between natural gas and oil prices, Energy Journal, 33 (2), 13-35.

Samuelson, P. A. (1970): The fundamental approximation theorem of portfolio analysis in terms of means, variances and higher moments, Review of Economic Studies, 37 (4), 537–542.

Satyanarayan, S. and Varangis, P. (1996): Diversification benefits of commodity assets in global portfolios, Journal of Investing 5(1), 69-78.

Sharpe, W. F. (964): Capital Asset Prices: A Theory of Market Equilibrium under Conditions of Risk, Journal of Finance, 19 (3), 425-42.

Silvennoinen, A. and Thorp, S. (2013): Financialization, crisis and commodity correlation dynamics, Journal of International Financial Markets, Institutions and Money, 24, 42-65.

Sklar, A. (1973): Random variables, joint distribution functions and copulas, Kybernetika, 9(6), 449–460.

Sortino, F.A., Van Der Meer, R. and Plantinga, A. (1999): The Dutch Triangle, Journal of Portfolio Management, 26(1), 50-58.

Steinbach, M. C. (2001): Markowitz Revisited: Mean-Variance Models in Financial Portfolio Analysis, SIAM Review, 43 (1), 31-85.

Tang, K. and Xiong, W. (2012): Index investment and financialization of commodities, Financial Analysts Journal, 68 (6), 54–74.

Unser, M. (2000): Lower Partial Moments as Measures of Perceived Risk: An Experimental Study, Journal of Economic Psychology, 21 (3), 253–280.

Villar, J.A. and Joutz, F.L. (2006): The Relationship between Crude Oil and Natural Gas Prices, Energy Information Administration, Office of Oil and Gas, Washington DC, October.

Zvi, B. and Rosansky, V. I. (1980): Risk and return in commodity futures, Financial Analysts Journal, 36 (3), 27-39.